# Metamaterial Model of Tachyonic Dark Energy

**Igor I. Smolyaninov**

Department of Electrical and Computer Engineering, University of Maryland, College Park, MD 20742, USA; E-Mail: smoly@umd.edu

**Abstract:** Dark energy with negative pressure and positive energy density is believed to be responsible for the accelerated expansion of the universe. Quite a few theoretical models of dark energy are based on tachyonic fields interacting with itself and normal (bradyonic) matter. Here we propose an experimental model of tachyonic dark energy based on hyperbolic metamaterials. Wave equation describing propagation of extraordinary light inside hyperbolic metamaterials exhibits 2+1 dimensional Lorentz symmetry. The role of time in the corresponding effective 3D Minkowski spacetime is played by the spatial coordinate aligned with the optical axis of the metamaterial. Nonlinear optical Kerr effect bends this spacetime resulting in effective gravitational force between extraordinary photons. We demonstrate that this model has a self-interacting tachyonic sector having negative effective pressure and positive effective energy density. Moreover, a composite multilayer SiC-Si hyperbolic metamaterial exhibits closely separated tachyonic and bradyonic sectors in the long wavelength infrared range. This system may be used as a laboratory model of inflation and late time acceleration of the universe.

**Keywords:** dark energy; analogue spacetime; hyperbolic metamaterial;

## 1. Introduction

Recent observational data have revealed accelerated expansion of the universe which cannot be explained by gravitational dynamics of ordinary matter. One of the possible explanations of these observations involves existence of a dark energy with negative pressure and positive energy density. It is assumed that gravitational repulsion due to dark energy accelerates present day expansion of the universe. Among various theoretical models of dark energy proposed so far, such as cosmological constant [1], quintessence [2], etc., tachyon-based models play a very prominent role [3-7]. A typically used relativistic tachyonic Lagrangian

$$L_T = -V(\phi)\sqrt{1-\partial_i\phi\partial^i\phi} \qquad (1)$$

indeed results in positive energy density

$$\rho = \frac{V(\phi)}{\sqrt{1-\partial_i\phi\partial^i\phi}} \qquad (2)$$

and negative pressure

$$P = -V(\phi)\sqrt{1-\partial_i\phi\partial^i\phi} < -\rho/3 \qquad (3)$$

necessary to explain the late time acceleration. However, physical origins of the tachyonic field and many issues related to dynamics of gravitationally interacting tachyons and normal (bradyonic) matter remain largely unclear. Motivated by these open questions and recent developments in electromagnetic metamaterials, we propose a hyperbolic metamaterial system which enables experimental study of gravitational dynamics of interacting tachyonic and bradyonic fields. Since this dynamics may have contributed to inflation and late time acceleration of our universe [3,5], such experiments may assist in refining theoretical models.

Our proposal is built on the recently developed hyperbolic metamaterial-based model of 2+1 dimensional gravity [8]. It appears that wave equation describing propagation of extraordinary light inside hyperbolic metamaterials exhibits 2+1 dimensional Lorentz symmetry [9]. The role of time in the corresponding effective 3D Minkowski spacetime is played by the spatial coordinate aligned with the optical axis of the metamaterial [10]. Nonlinear optical Kerr effect "bends" this spacetime resulting in effective gravitational interaction between extraordinary photons. In order for the effective gravitational constant to be positive, negative self-defocusing Kerr medium must be used [8]. These results are quite interesting given the fact that physical vacuum itself may behave as a hyperbolic metamaterial when subjected to very strong magnetic field [11,12]. Moreover, negative self-defocusing Kerr effect typically arises due to thermal expansion of the medium, which makes the effective 2+1 dimensional gravity a thermal effect obeying basic laws of thermodynamics. This feature of our model appears to be very attractive since recent theoretical developments strongly indicate thermodynamic origins of gravitational interaction [13,14].

## 2. Results and Discussion

Before we proceed to a model of interacting tachyonic fields, let us recall basic properties of hyperbolic metamaterials and their description using effective 2+1 dimensional Minkowski spacetime. Recent advances in electromagnetic metamaterials enable design of novel physical systems which can be described by effective space-times having very unusual metric and topological properties [15]. In particular, hyperbolic metamaterials (see Fig.1) offer an interesting experimental window into physics of Minkowski spacetimes, since propagation of extraordinary light inside a hyperbolic metamaterial is described by wave equation exhibiting 2+1 dimensional Lorentz symmetry. A detailed derivation of

**Figure 1.** Typical geometries of hyperbolic metamaterials: **(a)** metal wire array structure, and **(b)** multilayer metal-dielectric structure. The role of time in the effective 3D Minkowski spacetime is played by z coordinate aligned with the optical axis of the metamaterial, while $k_z$ behaves as an effective "energy". Depending on frequency range and materials used, both configurations may exhibit either bradyonic or tachyonic dispersion relations shown in **(c)**.

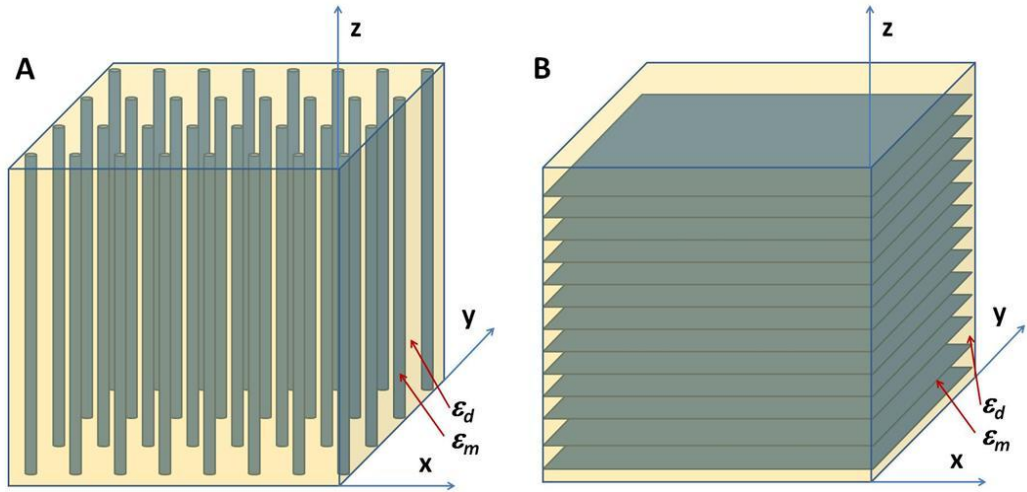

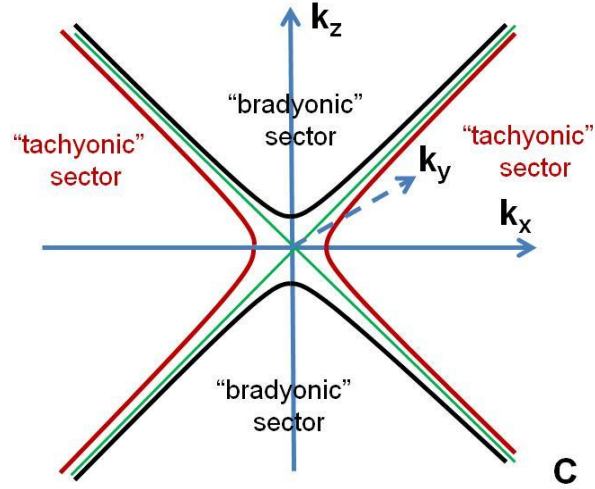

c

this result can be found in refs.[9,10]. Assuming that the metamaterial in question is uniaxial and non-magnetic, electromagnetic field inside the metamaterial may be separated into ordinary and extraordinary waves: vector $\vec{E}$ of the extraordinary light wave is parallel to the plane defined by the k–vector of the wave and the optical axis of the metamaterial. In the frequency domain (in some frequency band around $\omega=\omega_0$) the metamaterial may be described by anisotropic dielectric tensor having opposite signs of the diagonal components $\varepsilon_{xx}=\varepsilon_{yy}=\varepsilon_1$ and $\varepsilon_{zz}=\varepsilon_2$, while all the non-diagonal components are assumed to be zero in the linear optics limit. Propagation of extraordinary light in such a metamaterial may be described by a coordinate-dependent wave function $\varphi_\omega=E_z$ obeying the following wave equation [9,10]:

$$-\frac{\omega^2}{c^2}\varphi_\omega = \frac{\partial^2 \varphi_\omega}{\varepsilon_1 \partial z^2} + \frac{1}{\varepsilon_2}\left(\frac{\partial^2 \varphi_\omega}{\partial x^2} + \frac{\partial^2 \varphi_\omega}{\partial y^2}\right) \quad (4)$$

If $\varepsilon_1 > 0$ while $\varepsilon_2 < 0$, this wave equation coincides with the Klein-Gordon equation for a massive scalar field $\varphi_\omega$ in 3D Minkowski spacetime:

$$-\frac{\partial^2 \varphi_\omega}{\varepsilon_1 \partial z^2} + \frac{1}{(-\varepsilon_2)}\left(\frac{\partial^2 \varphi_\omega}{\partial x^2} + \frac{\partial^2 \varphi_\omega}{\partial y^2}\right) = \frac{\omega_0^2}{c^2}\varphi_\omega = \frac{m^{*2} c^2}{\hbar^2}\varphi_\omega \quad (5)$$

in which spatial coordinate $z=\tau$ behaves as a "timelike" variable. Eq.(5) describes world lines of massive particles obeying "relativistic" dispersion law

$$\frac{\omega^2}{c^2} = \frac{m^{*2} c^2}{\hbar^2} = \frac{k_z^2}{\varepsilon_1} - \frac{k_x^2 + k_y^2}{(-\varepsilon_2)}, \quad (6)$$

(see Fig.1(c)) which propagate in a flat 2+1 dimensional Minkowski spacetime. The wave vector component $k_z$ plays the role of effective energy, while vector $(k_x, k_y)$ plays the role of momentum. The effective mass squared $m^{*2}$ appears to be positive. Note that components of metamaterial dielectric tensor define the effective metric $g_{ik}$ of this spacetime: $g_{00}=-\varepsilon_1$ and $g_{11}=g_{22}=-\varepsilon_2$. This spacetime may be made "causal" by breaking the mirror and temporal symmetries of the metamaterial, which results in one-way light propagation along the timelike spatial coordinate [16], while "gravitational bending" of the effective spacetime may lead to an experimental model of the big bang [10]. In the weak gravitational field limit the effective Einstein equation

$$R_i^k = \frac{8\pi\gamma}{c^4}\left(T_i^k - \frac{1}{2}\delta_i^k T\right) \quad (7)$$

is reduced to

$$R_{00} = \frac{1}{c^2}\Delta\phi = \frac{1}{2}\Delta g_{00} = \frac{8\pi\gamma}{c^4}T_{00}, \quad (8)$$

where $\phi$ is the gravitational potential [17]. Since $z$ coordinate plays the role of time, while $g_{00}$ is identified with $-\varepsilon_1$, eq.(8) must be translated as

$$-\Delta^{(2)}\varepsilon_1 = \frac{16\pi\gamma^*}{c^4}T_{zz} = \frac{16\pi\gamma^*}{c^4}\sigma_{zz}, \quad (9)$$

where $\Delta^{(2)}$ is the 2D Laplacian operating in the $xy$ plane, $\gamma^*$ is the effective "gravitational constant", and $\sigma_{zz}$ is the $zz$ component of the Maxwell stress tensor of the electromagnetic field in the medium [18]:

$$\sigma_{zz} = \frac{1}{4\pi}\left(D_z E_z + H_z B_z - \frac{1}{2}\left(\vec{D}\vec{E}+\vec{H}\vec{B}\right)\right) \quad (10)$$

Detailed analysis performed in [8] indicates that nonlinear corrections to $\varepsilon_1$ due to Kerr effect lead to effective gravitational interaction between the extraordinary photons, and the sign of the third order nonlinear susceptibility $\chi^{(3)}$ of the hyperbolic metamaterial must be negative for the effective gravity to be attractive. It is also interesting to note that in the strong gravitational field limit this model contains 2+1 dimensional black hole analogs in the form of subwalength solitons [8].

Let us analyze how the basic framework outlined above can be extended to the tachyonic case. Very recently it has been noted [19] that depending on the frequency range and materials used, extraordinary photons in both hyperbolic metamaterial configurations shown in Fig.1(a,b) may exhibit a tachyonic dispersion relation. Let us analyze solutions of eq.(4) in the case where $\varepsilon_1 < 0$ while $\varepsilon_2 > 0$. It is clear that extraordinary photon propagation through such a metamaterial may still be described using an effective 2+1 dimensional Minkowski spacetime. However, the effective metric coefficients $g_{ik}$ of this spacetime change to $g_{00}=\varepsilon_1$ and $g_{11}=g_{22}=\varepsilon_2$, and the dispersion law of extraordinary photons changes to

$$-\frac{\omega^2}{c^2} = -\frac{\mu*^2 c^2}{\hbar^2} = \frac{k_z^2}{(-\varepsilon_1)} - \frac{k_x^2 + k_y^2}{\varepsilon_2}, \tag{11}$$

(see Fig.1(c)) where extraordinary photons acquire tachyonic imaginary effective mass $i\mu*$. It is easy to verify that eqs.(4,11) lead to correct sign of the trace of the effective 2+1 dimensional energy momentum tensor $T^{eff}_{ik}$ which coincides with the Maxwell stress tensor

$$T^{eff}_{ik} = \sigma_{ik} = \frac{1}{4\pi}\left(D_i E_k + H_i B_k - \frac{1}{2}\left(\vec{D}\vec{E} + \vec{H}\vec{B}\right)\right) \tag{12}$$

The contributions to $\sigma_{ik}$ which are made by a single extraordinary plane wave propagating inside the hyperbolic metamaterial may be calculated similar to [8]. Assuming without a loss of generality that the $B$ field of the wave is oriented along $y$ direction, the other field components may be found from Maxwell equations as

$$k_z B_y = \frac{\omega}{c}\varepsilon_1 E_x, \quad \text{and} \quad k_x B_y = -\frac{\omega}{c}\varepsilon_2 E_z \tag{13}$$

Taking into account the dispersion law (11) of the extraordinary wave, the contributions to $\sigma_{zz}$ and $\sigma_{xx}$ from a single plane wave are

$$\sigma_{zz} = -\frac{c^2 B^2 k_z^2}{4\pi\omega^2 \varepsilon_1} \quad \text{and} \quad \sigma_{xx} = -\frac{c^2 B^2 k_x^2}{4\pi\omega^2 \varepsilon_2}, \tag{14}$$

while $\sigma_{yy}=0$, leading to tachyonic negative $Tr\,T^{eff}=-B^2/4\pi$.

Similar to [8], nonlinear optical Kerr effect leads to gravity-like self-interaction of the tachyonic field. Taking into account that $g_{00}=\varepsilon_1$, the Einstein equation (8) translates into

$$\Delta^{(2)}\varepsilon_1 = \frac{16\pi\gamma*}{c^4}T_{zz} = \frac{16\pi\gamma*}{c^4}\sigma_{zz}, \tag{15}$$

where $\gamma*$ is the effective "gravitational constant". For a single plane wave eq.(15) may be rewritten as

$$\Delta^{(2)}\varepsilon_1 = \Delta^{(2)}\left(\varepsilon_1^{(0)} + \delta\varepsilon_1\right) = -4k_x^2\delta\varepsilon_1 = -\frac{4\gamma*B^2 k_z^2}{c^2\omega^2 \varepsilon_1}, \tag{16}$$

where the nonlinear corrections to $\varepsilon_1$ are assumed to be small, so that we can separate $\varepsilon_1$ into the constant background value $\varepsilon_1^{(0)}$ and weak nonlinear corrections (note that similar to [8], second order nonlinear susceptibilities $\chi^{(2)}_{ijl}$ of the metamaterial are assumed to be zero). These nonlinear corrections look like the Kerr effect assuming that the extraordinary photon wave vector components are large compared to $\omega/c$:

$$\delta\varepsilon_1 = \frac{\gamma*B^2 k_z^2}{c^2\omega^2 \varepsilon_1 k_x^2} \approx -\frac{\gamma*B^2}{c^2\omega^2 \varepsilon_2} = \chi^{(3)}B^2 \tag{17}$$

This assumption has to be the case if extraordinary photons may be considered as classic "particles". Eq.(17) establishes connection between the effective gravitational constant $\gamma^*$ and the third order nonlinear susceptibility $\chi^{(3)}$ of the hyperbolic metamaterial. Similar to the "bradyonic case" considered in [8], the sign of $\chi^{(3)}$ must be negative for the effective gravity to be attractive (since $\varepsilon_2>0$). Since most liquids exhibit large and negative thermo-optic coefficient resulting in large and negative $\chi^{(3)}$, and there exist readily available ferrofluid-based hyperbolic metamaterials [20], laboratory experiments with gravitationally self-interacting tachyonic fields appear to be realistic in the near future. Moreover, as we will demonstrate below, even more curious case of coexisting mutually-interacting tachyonic and bradyonic fields seems to be no more difficult to realize. Since gravitational dynamics of mutually interacting tachyonic and bradyonic fields may have contributed to inflation and late time acceleration of our universe [3,5], such experiments would be very interesting.

Let us consider a multilayer metal-dielectric metamaterial shown in Fig.1(b). The diagonal components of its dielectric tensor may be obtained using Maxwell-Garnett approximation [21]:

$$\varepsilon_1 = \varepsilon_{xy} = n\varepsilon_m + (1-n)\varepsilon_d \qquad (18)$$

$$\varepsilon_2 = \varepsilon_z = \frac{\varepsilon_m \varepsilon_d}{(1-n)\varepsilon_m + n\varepsilon_d} \qquad (19)$$

where $n$ is the volume fraction of the metallic phase (assumed to be small), and $\varepsilon_m$ and $\varepsilon_d$ are the dielectric permittivities of the metal and dielectric phase, respectively. A suitable choice of $n$ is known to lead to hyperbolic behaviour in such metamaterials [21]. However, ordinary metals cannot be used in a hyperbolic metamaterial design having closely spaced tachyonic and bradyonic frequency bands due to their broadband metallic ($\varepsilon_m<0$) behaviour. This difficulty may be overcome by using such materials as SiC, which have narrow Restsrahlen metallic bands in the long wavelength infrared (LWIR) range. Silicon carbide may be used in combination with such material as Si, which exhibits broadband dielectric behaviour in the LWIR having $\varepsilon_d\sim10.9$. Indeed, our calculations presented in Fig.2 indicate that around n~0.3 a multilayer SiC-Si metamaterial does have pronounced closely spaced tachyonic and bradyonic frequency bands, while having relatively low losses. The calculated values of $\varepsilon_1$ and $\varepsilon_2$ are based on the measured optical properties of SiC reported in [22]. A somewhat similar result has been reported in [21] for a SiC-Si wire array structure, but unlike our multilayer design, fabrication of SiC-Si wire arrays may prove extremely difficult. We should also point out that such materials as $SiO_2$, $Al_2O_3$, etc. also have pronounced Restsrahlen bands in the LWIR range, so that they can be used in the layered design with equal success. If native undoped Si is used in the metamaterial design, its nonlinearity will be dominated by negative thermo-optic coefficient due to thermal expansion. Thus, the desired negative $\chi^{(3)}$ metamaterial will be obtained resulting in positive $\gamma^*$.

We should also point out that $\varepsilon$ and $\chi^{(3)}$ tensors of the metamaterial do not need to stay coordinate independent. Spatial behaviour of the dielectric permittivity tensor components may be engineered so that the background metric may closely emulate metric of the universe during inflation [23]. On the other hand, engineered higher order nonlinear susceptibility terms $\chi^{(n)}$ may be used to emulate the desired functional form of the tachyonic potential $V(\phi)$ (see eqs.(1-3)). As a result, various scenarios of tachyonic inflation [3] will become amenable to direct experimental testing.

**Figure 2.** (a) Calculated diagonal components of the dielectric permittivity tensor of a multilayer SiC-Si metamaterial in the LWIR frequency range. The volume fraction of SiC equals n=0.3. Two closely spaced tachyonic and bradyonic hyperbolic bands arise near $\lambda$=11µm. (b) Calculated $Im(\varepsilon)/|\varepsilon|$ for both directions indicate low loss hyperbolic behavior inside the bands.

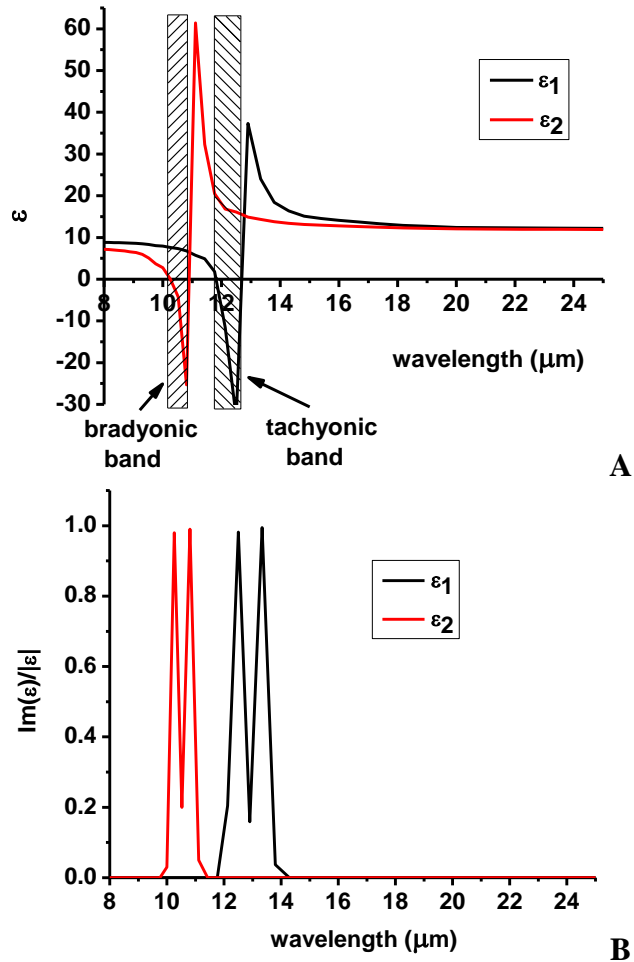

## 3. Conclusions

In conclusion, we have demonstrated that extraordinary photons in a composite multilayer SiC-Si hyperbolic metamaterial exhibit closely separated tachyonic and bradyonic frequency bands around λ=11μm. Nonlinear optical Kerr effect leads to effective gravitational interaction of photons in these bands. This interaction may be used to study gravitational dynamics of tachyonic and bradyonic fields, which is responsible for inflation and late time acceleration of the universe in the tachyonic models of dark energy. While metamaterial losses constitute an important performance-limiting issue for this model, loss compensation using gain media [24] is known to be able to overcome this problem. In our particular case loss compensation is simplified by the fact that in our metamaterial design the tachyonic and the bradyonic bands are located very close to each other.

**Conflicts of Interest**

The author declares no conflict of interest.